\def\br{{\bf r}}

\def\bk{{\bf k}}
\def\bK{{\bf K}}
\def\bG{{\bf G}}
\def\bq{{\bf q}}

\documentstyle[aps,prb,eqsecnum,twocolumn,epsfig]{revtex}

\begin{document}

\draft

\title{Enhancements to the {\it GW} space-time method}

\author{L. Steinbeck$^\dagger$\footnotemark[1],
          A. Rubio$^\ddagger$, L. Reining$^\S$, M. Torrent$^\S$\cite{foot0}, 
          I. D. White$^+$, and R. W. Godby$^\dagger$}
\address{$^\dagger$Department of Physics, University of York, Heslington,
            York YO1 5DD, UK}
\address{$^\ddagger$ Departamento Fisica Teorica, Universidad de Valladolid,
            E-47011 Valladolid, Spain }
\address{$^\S$ Laboratoire des Solides Irradi\'es, UMR 7642 CNRS - CEA/CEREM,
         Ecole Polytechnique, Palaiseau, F-91128, France }
\address{$^+$Cavendish Laboratory, University of Cambridge, Madingley Road,
             Cambridge, CB3 0HE, UK}

\date{\today}

\maketitle

\begin{abstract}
We describe the following new features which significantly enhance the power 
of the recently developed real-space imaginary-time $GW$ scheme (Rieger 
{\it et al.}, Comp.~Phys.~Commun. {\bf 117}, 211 (1999)) for the calculation of 
self-energies and related quantities of solids:
(i) to fit the smoothly decaying time/energy tails of the dynamically 
screened Coulomb interaction and other quantities to model functions, 
treating only the remaining time/energy region close to zero numerically 
and performing the Fourier transformation from time to energy and vice 
versa by a combination of analytic integration of the tails 
and Gauss-Legendre quadrature of the remaining part and
(ii) to accelerate the convergence of the band sum in the calculation of the
Green's function by replacing higher unoccupied eigenstates by free 
electron states (plane waves).
These improvements make the calculation of larger systems (surfaces, clusters,
defects etc.) accessible.

\end{abstract}


\section{Introduction} 
\label{sec:intro}

Density-functional calculations provide reliable information about the ground
state properties of electron systems but give, in principle, no access to
the excitation spectrum of the system under study. Excitations can be described 
by many-body perturbation theory which is, however, at present only
computationally feasible for real materials in its simplest form, the $GW$
approximation of Hedin.\cite{Hedin65,Hedlu69} The latter gives a 
comparatively simple expression for the self-energy operator, which allows
the one-particle Green's function of an interacting many-electron system
to be described in terms of the Green's function of a hypothetical
non-interacting system with an effective potential. The Green's function
contains information not only about the ground-state density and energy
but also about the quasiparticle (QP) spectrum. The $GW$ approximation has
been successfully applied to the calculation of QP bandstructures of 
semiconductors and other materials,\cite{HYL286,GSS88,NHL89,Aryasetiawan92} 
for a recent review see Ref.\ \onlinecite{AryaGun98}.
\renewcommand{\thefootnote}{\fnsymbol{footnote}}
\footnotetext[1]{present address: Institut f\"ur Festk\"orper- und 
           Werkstofforschung e.V., Postfach 270016, 01171 Dresden, Germany, 
           email: lutz@tmps16.mpg.tu-dresden.de, Tel.: (+49) 351 463 4608, 
           Fax:  (+49) 351 463 7029}

The real-space imaginary-time $GW$ method, first proposed by Rojas 
{\it et al.}\cite{RGN95} and -- in a revised form -- described 
in detail by Rieger {\it et al.}\cite{RSWRG99} (we will refer to this 
paper as CPC I in the following) offers a more favourable scaling of the
computational effort with system size than conventional reciprocal-space 
$GW$ schemes.\cite{RSWRG99} 
It substantially reduces the computational effort and allows to study
larger systems than previously possible without resorting to further
approximations such as plasmon-pole models\cite{HYL286} for the energy
dependence of the screened interaction or model dielectric 
functions.\cite{LEL82}

The new features outlined in the present paper, particularly the new
treatment of the (imaginary) time/energy dependence, further reduce the 
computational effort
of the space-time $GW$ scheme by almost an order of magnitude. This is
achieved by fitting the smoothly decaying large energy/time tails of
all quantities involved in a $GW$ calculation to simple model functions
and treating the remaining time/energy region numerically on a
Gauss-Legendre grid rather than using an equidistant grid and
fast Fourier transformations (FFT) from time to energy and vice versa.
In the new scheme these Fourier transformations are performed by
a combination of analytic integration of the tails and
Gauss-Legendre quadrature of the remaining part.
Another improvement of the method concerns the convergence of the 
calculated Green's function with the number of unoccupied eigenstates
entering the eigenstate (band) sum in the Green's function Eq.\ (\ref{GLDAST})
below. Higher unoccupied eigenstates are approximated by plane waves. This 
considerably reduces the number of eigenstates and energies
which have to be computed in a density-functional calculation (usually
within the local density approximation (LDA)) preceding a calculation
of the self-energy with a given accuracy.

The present paper is organized as follows: first we give a brief summary
of the real-space imaginary-time $GW$ scheme in order to clarify notation
in reference to CPC I (Section \ref{sec:method}). Then we describe the 
new treatment of the time/energy dependence (Section \ref{sec:glg}) and the 
plane-wave substitution for accelerating the unoccupied-state sum convergence 
of the Green's function (Section \ref{sec:pwt}).

\section{Summary of the real-space imaginary-time {\it GW} method}
\label{sec:method}

In the real-space imaginary-time $GW$ method\cite{RGN95,RSWRG99} for computing 
electron self-energies and related quantities such as dielectric
response functions and quasiparticle energies the basic quantities Green's 
function, dielectric response function, dynamically screened Coulomb interaction 
and self-energy are represented on a real-space grid and on the 
imaginary time axis. In those intermediate steps of the calculation where it
is computationally more efficient to work in reciprocal space and imaginary
energy we change to the latter representation by means of Fourier transforms.
The choice of representing the time/energy dependence on
the imaginary instead of on the real axis allows us to deal with smooth,
decaying quantities which give faster convergence. To obtain the self-energy
eventually on the real energy axis, we fit a model function to the computed
self-energy on the imaginary axis, and continue it analytically to the real
axis. The energy dependence of the dynamically screened interaction is fully 
taken into account within the method. The computational effort 
scales quadratically with the number of atoms in the unit cell
and linearly with the number of energy points $N_{\omega}$ used to represent 
the energy dependence.\cite{RSWRG99}  

First, the zeroth-order Green's function is constructed in real space and 
imaginary time:
\begin{eqnarray}
\label{GLDAST}
\lefteqn{
G_{LDA}(\br,\br';i\tau) } && \\[12pt]
 && = \left\{\begin{array}{ll}
 \phantom{-}
 i \sum\limits_{n\bk}^{occ}\Psi_{n\bk}(\br)\Psi^*_{n\bk}(\br')
   \exp(\epsilon_{n\bk}\tau) , & \tau > 0, \\[12pt]
-i \sum\limits_{n\bk}^{unocc}\Psi_{n\bk}(\br)\Psi^*_{n\bk}(\br')
   \exp(\epsilon_{n\bk}\tau) , & \tau < 0, \nonumber \\ \\
\end{array}
\right.
\end{eqnarray}
from the LDA wavefunctions $\Psi_{n\bk}(\br)$ and eigenvalues
$\epsilon_{n\bk}$.
Then the RPA irreducible polarizability is formed in real space and imaginary 
time:
\begin{equation}
\label{chi0}
\chi^0(\br,\br';i\tau) = -iG_{LDA}(\br,\br';i\tau)G_{LDA}(\br',\br;-i\tau),
\end{equation}
and Fourier transformed to reciprocal space and imaginary energy and
the symmetrised dielectric matrix\cite{BAR86} is constructed in reciprocal
space,
\begin{equation}
\label{symdielm}
\tilde\epsilon_{\bG\bG'}(\bk,i\omega) = \delta_{\bG\bG'}  
 {}-\frac{4\pi}{\left|\bk+\bG\right|\left|\bk+\bG'\right|}
      \chi^0_{\bG\bG'}(\bk,i\omega).
\end{equation}
After that the symmetrised dielectric matrix is inverted for each $\bk$ 
point and each imaginary energy in reciprocal space and the
screened Coulomb interaction is calculated:
\begin{equation}
W_{\bG\bG'}(\bk,i\omega)  = 
\frac{4\pi}{\left|\bk+\bG\right|\left|\bk+\bG'\right|}
\tilde\epsilon^{-1}_{\bG\bG'}(\bk,i\omega),
\end{equation}
and Fourier transformed to real space and imaginary time.
From that the self-energy operator
\begin{equation}
\Sigma(\br,\br';i\tau) = iG_{LDA}(\br,\br';i\tau)W(\br,\br';i\tau),
\end{equation}
and its expectation values $\left<\bq n|\Sigma(i\tau)|\bq n\right>$
are computed. The latter are Fourier transformed to imaginary energy and
fitted to a model function allowing analytic continuation onto the real 
energy axis and evaluation of the quasiparticle corrections to the LDA 
eigenvalues by first-order perturbation theory in 
$\left\langle \Sigma -V_{\text{xc}}^{\text{LDA}}\right\rangle $.
Since all quantities go to zero with increasing $|\br - \br'|$ we use a finite
cutoff region in real space which we call the interaction cell.
Further details of the method can be found in CPC I.

\section{New treatment of time/energy dependence}
\label{sec:glg}

\subsection{motivation and basic idea}

\begin{figure}
\epsfxsize=3.25in \epsfbox{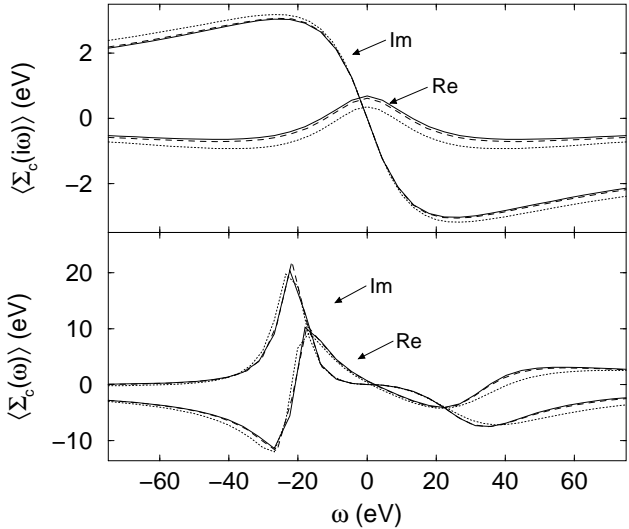}
\caption{Convergence of the matrix element of the correlation self-energy 
$\langle \Sigma_c(i\omega) \rangle$ for the valence-band top 
$\Gamma_{25\prime}^v$
of silicon with respect to $\omega_{max}$ with a fixed energy grid spacing of
$\Delta\omega$= 0.17 Har. The top panel shows the calculated self-energy on
the imaginary axis with the analytically continued dependence on real energy
shown in the lower panel. The lines correspond to  $\omega_{max}$= 5 Har
(dotted), 10 Har (dashed) and 20 Har (solid), i.~e.~30, 60, and 120 FFT grid
points, respectively. }
\label{sigwmax}
\end{figure}

The functions we are dealing with are relatively smooth on the imaginary
time/energy axis. This allows to employ a regular time/energy grid which has the
advantage that the Fourier transformation from imaginary time to imaginary
energy and vice versa can be done efficiently by fast Fourier
transformation (FFT). However, we still need of the order of 100 grid 
points\cite{foot1}
for good convergence (resulting quasiparticle energies converged within 30
meV with respect to the $\tau$/$\omega$ grid parameters). This point is
illustrated by Figure \ref{sigwmax} showing the matrix
element of the correlation self-energy\cite{foot2} for the uppermost valence
band of Si at $\Gamma$, calculated (with $\Delta\omega$ = 0.17 Hartree) with 
30, 60, and 120 FFT grid points (a) on the imaginary energy axis and (b) 
analytically continued to real energies. 
Crucially, the convergence on the imaginary axis transforms into a
convergence of similar quality on the real axis upon analytic continuation.
Looking at the time/energy behaviour of the key quantities, 
particularly those which have to be Fourier transformed such as polarizability,
screened interaction and the matrix elements of the self-energy (see Figure
\ref{siform}) we observe that they possess nontrivial structure only in 
the region close to $i\tau = 0$  ($i\omega = 0$) whereas they decay 
smoothly to zero for large
imaginary times or energies. The FFT grid has to be large enough to take
account of the tails (reduce aliasing) and at the same time it needs
to be sufficiently dense to describe the structure in the region close to 
the origin properly.

This suggests another approach: represent the functions on a suitably
chosen grid in a fixed and comparatively small time/energy interval and fit
the large $i\tau$/$i\omega$ tails to simple model functions which can be
Fourier transformed analytically, a method suggested by Blase {\it et
al.}\cite{Blase99} in the context of their earlier mixed-space 
method.\cite{Blase95} For the part handled numerically we choose a
Gauss-Legendre (GL) grid (linearly transformed from (-1,1) to 
(0,$\tau_{max}$) or (0,$\omega_{max}$), respectively). This turns out to 
be very efficient since the
functions have to be computed for a relatively small number of time or energy
points only and this computation of the functions is much more time-consuming
than the Fourier transformations themselves which are done by Gaussian
quadrature over the numerical values and adding the Fourier transform of the
tail.\cite{foot3} The fit of the tail only needs to be performed 
whenever a quantity has to be Fourier
transformed from $i\tau$ to $i\omega$ or vice versa. The rest of the
calculation is restricted to the GL grid.
Hence the following quantities have to be fitted:
(1) the polarizability $\chi^0_{{\bf GG}'} ({\bf k}, i\tau)$,
(2) the screened Coulomb interaction $W_{{\bf GG}'} ({\bf k}, i\omega)$
and
(3) the matrix elements of the correlation part of the self-energy
$\langle \bq n|\Sigma_c (i\tau)|\bq n \rangle$.

The advantages of this treatment of the time/energy dependence are
obvious: all quantities have to be computed for a much smaller number
of imaginary times/energies thus saving computational 
time and reducing storage requirements while retaining the flexibility 
to accomodate general functional forms of the energy dependence and not 
being restricted to particular forms such as plasmon-pole models.

\subsection{Polarizability}

The imaginary-time dependence of the Green's function derives from the
the decaying exponentials $exp(\epsilon_{n{\bf k}}\tau)$ with
$\epsilon_{n{\bf k}}<0\; (>0)$ for $\tau>0\; (<0)$ 
in Eq.\ (\ref{GLDAST}). The asymptotic behavior at large imaginary times is
determined by the slowest-decaying exponentials and can thus be approximated by
a single exponential. This asymptotic imaginary-time dependence carries over
to the polarizability Eq.\ (\ref{chi0}). For that reason we fit the
large-$i\tau$ tails of each $({\bf GG}' {\bf k})$ element of the 
polarizability $\chi^0_{{\bf GG}'} ({\bf k}, i\tau)$ \cite{foot4} 
to a decaying exponential $a\; exp(-b\tau)$ 
(with $b>0$ and for $\tau \ge 0$ only since $\chi^0$ is symmetric in $\tau$). 
The two fit parameters $a$ and $b$ are exactly
determined by fitting two time points: the outermost GL grid
point and one additional point at $1.3 \tau_{max}$. This fitting
procedure turns out to be very reliable.

The Fourier transformation from $i\tau$ to $i\omega$ is done in the following
way:
\begin{eqnarray}
\label{polfit}
\chi^0(i\omega_j) &=& 
\sum_{i=-i_{max}}^{i_{max}} p_i\; [{\bar \chi}^0 
(i\tau_i) - a\; exp(-b|\tau_i|) ] exp(-\omega_j \tau_i)  \nonumber \\
&&+ \int\limits_{-\infty}^{\infty}d\tau\; a\; exp(-b|\tau|) 
exp(-\omega_j \tau) \nonumber \\
&=& 2 \sum_{i=1}^{i_{max}} p_i\; 
[{\bar  \chi}^0 (i\tau_i) - a\; exp(-b|\tau_i|) ]\;
cos (\omega_j \tau_i) \nonumber \\
&&+ \frac{2 a}{b^2 + \omega^2_j}
\end{eqnarray}
with GL grid points $\tau_i$ and $\omega_j$, GL weights $p_i$, fit parameters
$a$ and $b$, and ${\bar \chi}^0 (i\tau) = -i \chi^0_{{\bf GG}'} ({\bf k},
i\tau)$. 

For a small number of matrix elements $\chi^0_{{\bf GG}'} ({\bf k},
i\tau)$ (typically less than 5\% of all matrix elements as long as $\tau_{max}$ 
is large enough to accomodate all the nontrivial structure of $\chi_0 (i\tau)$)
the large $i\tau$ tails cannot be fitted
to a decaying exponential because they increase or change sign. This is only
the case for small matrix elements where the function is already close to
zero at $\tau_{max}$ anyway. We set $\chi^0 (1.3\tau_{max})$ to
$0.1 \chi_0 (\tau_{max})$ there, i.~e.~choose a reasonable decaying constant
which takes the (already small) function smoothly to zero. 
Simply setting the matrix element to zero for $\tau > \tau_{max}$ would render 
the ensuing fit of $W_{{\bf GG}'} ({\bf k}, i\omega)$ unneccessarily difficult.
Anyhow, $\tau_{max}$  is a convergence parameter which can be varied to check the 
quality of the results.

\subsection{Dynamically screened Coulomb interaction}

The large-imaginary-energy tail of the dynamically screened interaction 
$W_{{\bf GG}'} ({\bf k},i\omega)$ is fitted to the Fourier transform of a
decaying exponential
\begin{equation}
\int\limits_{-\infty}^{\infty}d\tau\; a\; exp(-b|\tau|) exp(\pm i\omega\tau)
= \frac{2 a b}{b^2 + \omega^2} = \frac{\alpha}{\beta^2 + \omega^2}.
\end{equation}
The energy region where $W$ is treated numerically has to be large enough to
comprise the nontrivial structure of $W(i\omega)$. We found that
$\omega_{max}$ should be between 3 and 10 times the plasmon energy\cite{foot5}
for good convergence.
We could perform the tail fit along similar lines as that of $\chi_0$,
i.~e.~subtract the analytic tail function from the given imaginary-energy
W in (0, $\omega_{max}$), Fourier transform this difference numerically
and add the analytically given Fourier transform of the function
fitted to the tail back in. However, for a large number of matrix elements
$W_{{\bf GG}'} ({\bf k},i\omega)$ the tail fit yields a negative $\beta^2$
because they decay more rapidly than $1/\omega^2$. In this case the
function $\alpha/(\beta^2 + \omega^2)$ has a pole inside the interval
(0, $\omega_{max}$) which does not allow the analytic Fourier transformation 
to be performed and which makes the numerical Fourier transformation of the
difference between W and the fit function virtually impossible to compute.
That is why we integrate the analytic tail function from $\omega_{max}$ to
$\infty$ in this case, provided that $\beta^2 > - \omega^2_{max}$. Part of this
integral can still be solved analytically whereas the remainder is treated
numerically. The Fourier transform 
${\bar W} (i\tau) = -i W_{{\bf GG}'} ({\bf k},i\tau)$ is given by:
\begin{eqnarray}
\label{wfit}
{\bar W} (i\tau_i) &=& \frac{1}{\pi} \sum_{j=1}^{j_{max}} p_j\; W (i\omega_j)\;
cos(\omega_j\tau_i) \nonumber \\
&&+ \frac{1}{\pi} \int\limits_{\omega_{max}}^{\infty}
d\omega \frac{\alpha}{\beta^2 + \omega^2}\; cos(\omega\tau_i)
\end{eqnarray}
with
\begin{eqnarray}
\label{wfit2}
\lefteqn{
\frac{1}{\pi}\int\limits_{\omega_{max}}^{\infty}d\omega 
\frac{\alpha}{\beta^2 + \omega^2}\; cos(\omega\tau) =} && \nonumber \\
&&- \frac{\alpha}{\pi} \int\limits_{\omega_{max}}^{\infty} d\omega
\frac{\beta^2}{\beta^2 + \omega^2} \frac{cos(\omega\tau)}{\omega^2} 
\nonumber \\
&&+\frac{\alpha}{\pi} \left[ \frac{cos(\omega_{max}\tau)}{\omega_{max}} - 
\tau Si (\omega_{max} \tau)\right]
\end{eqnarray}
with the sine integral
\begin{equation}
Si(\omega_{max}\tau) = \int\limits_{\omega_{max}}^{\infty} d\omega
\frac{sin(\omega\tau)}{\tau}.
\end{equation}
Here $\tau_i$ and $\omega_j$ are the GL grid points, $p_j$ the GL weights,
$\alpha$ and $\beta^2$ the fit parameters and $\omega_{j_{max}}$
the outermost GL grid point. The integral on the right hand side of
Eq.\ (\ref{wfit2}) is solved
numerically using a transformed GL grid. It converges rapidly since the integrand
is going to zero like $1/\omega^4$. The second part of Eq.\ (\ref{wfit2}) is
given analytically. In this way most $W_{{\bf GG}'} ({\bf k},i\omega)$ 
can be fitted except for a small number where $\beta^2 \le - \omega^2_{max}$ 
(this only occurs for matrix elements which are small anyway). In the latter 
case we take the correct value at $\omega_{max}$ smoothly to zero by setting 
$\beta^2$ to $-0.9\omega^2_{max}$. Again, the quality of the results can be
checked by varying $\omega_{max}$.

\subsection{Matrix elements of correlation self-energy}

The matrix elements of the correlation self-energy
are fitted in a similar way as the polarizability. As
the asymptotic time dependence of the self-energy is determined by that of the
Green's function which is `shorter-ranged' (in imaginary time) than $W$ 
we again fit to a
decaying exponential $a\;exp(-b\tau)$. This time, however, we have to perform
separate fits on the positive and negative half-axis since the self-energy
is not symmetric in imaginary time. Therefore we obtain two contributions
to $\langle \Sigma_c (i\omega) \rangle$ for positive imaginary energies
(with ${\bar \Sigma_c} (i\tau) = -i \langle\Sigma_c(i\tau)\rangle$):
\begin{eqnarray}
\label{sigmafit}
\langle \Sigma_c (i\omega_j) \rangle &=& \sum_{i=1}^{i_{max}} p_i\;
[{\bar \Sigma}_c (i\tau_i) - a_+ exp(-b_+ |\tau_i|)] \nonumber \\
& & \times\;exp(-i\omega_j\tau_i) + \frac{a_+}{b_+ + i \omega_j} \nonumber \\
&+& \sum_{i=-1}^{-i_{max}} p_i\; [{\bar \Sigma}_c (i\tau_i) - a_- 
exp(-b_- |\tau_i|)] \nonumber \\
& & \times\;exp(-i\omega_j\tau_i) + \frac{a_-}{b_- + i \omega_j} 
\end{eqnarray}
here $\tau_i$ and $\omega_j$ denote GL grid points, $p_i$ GL 
weigths, and $a_+$, $a_-$, $b_+$, and $b_-$ are the fit parameters. 
The self-energy matrix elements on the
negative imaginary-energy half-axis are given by 
$\langle \Sigma_c (-i\omega) \rangle = \langle \Sigma_c (i\omega) \rangle^* $.

All matrix elements of $\Sigma_c$ could be fitted in this way for all the
systems investigated so far (Si, Ge, GaN).

\subsection{Tests for bulk silicon and zincblende GaN}
\label{sec:testsi}

In order to test the stability and accuracy of the tail fit and the
convergence with the GL grid size we performed self-energy calculations
for bulk Si (cutoff parameters are given in Table \ref{glgpar} and
further details are as in CPC I).
\begin{table}
\caption{Cutoff and grid parameters used for the test calculations for
Si and zincblende GaN, respectively, in section \ref{sec:glg} of the present 
work.}
\begin{tabular}{lcc}
 parameter & Si & GaN   \\
\tableline
 LDA plane-wave cutoff (in Ry)                          & 13.5 & 50. \\
 $GW$ plane-wave cutoff\tablenotemark[1]  (in Ry)       &  19. & 38. \\
 $GW$ real-space grid       & $9\times9\times9$  & $12\times12\times12$ \\
 band cutoff (in Ry)                                    &  10. & 16.  \\
 size of {\bf k} grid          & $4\times4\times4$ & $4\times4\times4$   \\
 range $\tau_{max}$ of time grid\tablenotemark[2] (in a.u.) & 6. & 6.  \\
 size of time (energy) grid\tablenotemark[2]                & 15 & 15 \\
\end{tabular}
\label{glgpar}
\tablenotetext[1]{Energy cutoff corresponding to radius of circumscribing 
sphere, see Ref.\ \onlinecite{RSWRG99}.} 
\tablenotetext[2]{This parameter is varied in the tests of Section 
\ref{sec:glg}.}
\end{table}

\begin{figure}
\epsfxsize=3.25in \epsfbox{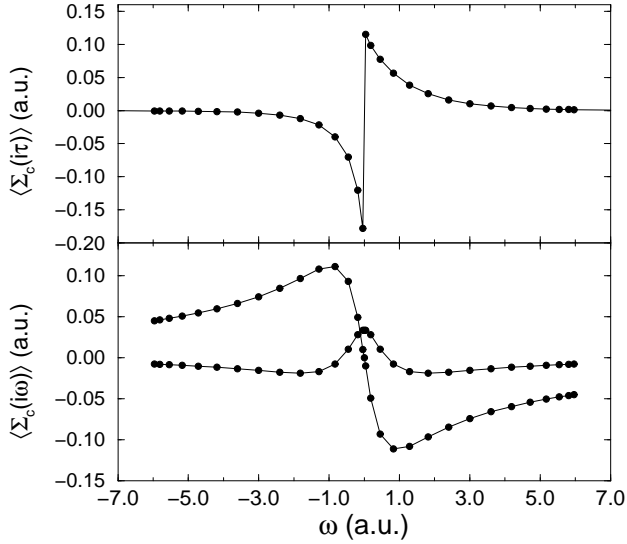}
\caption{Matrix elements $\langle \Sigma_c(i\tau) \rangle$ (top panel) and 
$\langle \Sigma_c(i\omega) \rangle$ (lower panel) of the correlation 
self-energy for the valence-band top $\Gamma_{25\prime}^v$ of silicon. 
The Gauss-Legendre grid points for 
$\tau_{max}$ = $\omega_{max}$ = 6 a.u. and $N_{\tau}$ = $N_{\omega}$ = 15 are
indicated by filled circles. This figure gives an idea about a typical 
functional form on the imaginary time (energy) axis and a typical 
Gauss-Legendre grid which has to be used to Fourier transform this function 
with good convergence. The other quantities
which have to be Fourier transformed from time to energy or vice versa have
similar (or less) structure.}
\label{siform}
\end{figure}
Figure \ref{siform} gives an idea of the sampling of the matrix element of
the correlation self-energy for the uppermost valence band at $\Gamma$ on
the imaginary time and imaginary energy axis by a GL grid with $\tau_{max}$ =
$\omega_{max}$ = 6 a.u. and $N_{\tau}$ = $N_{\omega}$ = 15. The convergence of
two typical quasiparticle energies with the number of GL grid points and the
size of the imaginary time/energy region treated explicitly is summarized in
Table \ref{siconv}.\cite{foot6}
We chose the lowest conduction state $\Gamma_{15}^{c\prime}$ 
and the valence-band bottom $\Gamma^v_1$. The latter is in our experience the 
slowest-converging quasiparticle energy which is most sensitive to details of
the calculation and cutoff parameters. Obviously, $N_{\tau}$ = $N_{\omega}$
and $\tau_{max}$ = $\omega_{max}$ have to be increased simultaneously 
since the main structure of the functions we are dealing with is restricted to
a limited region around the origin. We observe that 15 GL grid points and 
$\tau_{max}$ = $\omega_{max}$ = 5 a.u. are sufficient to converge even the
slowest-converging valence-band bottom to within 30 meV. The same accuracy is
obtained with FFT grids for $\tau_{max}$ = $\omega_{max}$ = 20 a.u. and
120 grid points, resulting in QP energies of 3.22 eV ($\Gamma_{15}^{c\prime}$)
and $-11.58$ eV ($\Gamma^v_1$).

Our $GW$ calculations for GaN in the zincblende structure were
carried out at the experimental lattice constant (a = 8.54 a.u.). 
The LDA wavefunctions and eigenvalues used in the self-energy calculation were
obtained from a standard plane-wave pseudopotential calculation. The Ga 4s and
4p states and the N 2s and 2p states were treated as valence states and the
soft  pseudopotentials of Troullier and Martins \cite{TM91} were used.
The cutoff parameters are given in Table \ref{glgpar}. 
Table \ref{ganconv} shows the convergence of the resulting QP energies 
at $\Gamma$ and $X$ as a function of the time/energy GL grid parameters.
From these results we conclude that the speed of convergence of the QP 
energies of GaN with respect to the imaginary time/energy grid is similar 
to that found for bulk Si.
\begin{table}
\caption{Convergence of quasiparticle energies $\Gamma_{15}^{c\prime}$ and
$\Gamma_1^v$ (in eV, top of valence band has been set to zero) for Si with 
respect to Gauss-Legendre grid region $\tau_{max} = \omega_{max}$ (in a.u.) 
and number of grid points $N_{\tau} = N_{\omega}$.}
\begin{tabular}{lcccccc}
$\tau_{max} = \omega_{max}$ & \multicolumn{6}{c}{$N_{\tau} = N_{\omega}$} \\
        &  10  &  12  &  15  &  18  &  21  &  25 \\
\tableline
$\Gamma_{15}^{c\prime}$ &&&&&& \\
\tableline
    3.  & 3.27 & 3.28 & 3.28 &      &      &     \\
    4.  & 3.24 & 3.24 & 3.24 &      &      &     \\
    5.  &      & 3.23 & 3.23 & 3.23 &      &     \\
    6.  &      &      & 3.23 & 3.23 & 3.22 &     \\
    7.  &      &      &      & 3.22 & 3.23 & 3.22 \\
\tableline
$\Gamma_1^v$ &&&&&& \\
\tableline
    3.  & -11.54 & -11.53 & -11.52 &        &        &        \\
    4.  & -11.58 & -11.52 & -11.52 &        &        &        \\
    5.  &        & -11.53 & -11.56 & -11.56 &        &        \\
    6.  &        &        & -11.56 & -11.58 & -11.60 &        \\
    7.  &        &        &        & -11.63 & -11.58 & -11.59 \\
\end{tabular}
\label{siconv}
\end{table}
\begin{table}
\caption{Convergence of quasiparticle energies (in eV, top of
valence band has been set to zero) at the $\Gamma$ 
and $X$ point of zincblende GaN with respect to Gauss-Legendre grid 
region $\tau_{max} = \omega_{max}$ (in a.u.) and number of 
time/energy grid points $N_{\tau} = N_{\omega}$. }
\begin{tabular}{lcccc}
$N_{\tau} = N_{\omega}$ &    12  & 15    & 18    & 25 \\
$\tau_{max} = \omega_{max}$ &    4.  & 5.    & 6.    &  8.\\
\tableline
$\Gamma_1^v$            & -15.88 &-15.95 &-15.93 &-15.94 \\ 
$\Gamma_1^c$            &   3.03 &  3.00 &  3.00 &  2.99 \\
$\Gamma_{15}^c$         &  11.56 & 11.54 & 11.53 & 11.52 \\[1mm]
$X_1^v$                 & -12.96 &-13.00 &-12.97 &-12.98 \\
$X_3^v$                 &  -6.38 & -6.39 & -6.38 & -6.38 \\
$X_5^v$                 &  -2.66 & -2.66 & -2.66 & -2.66 \\
$X_1^c$                 &   4.43 &  4.41 &  4.40 &  4.39 \\
$X_1^{c\prime}$         &   7.91 &  7.88 &  7.87 &  7.86 \\
$X_3^c$                 &  13.23 & 13.19 & 13.16 & 13.14 \\
$X_5^c$                 &  15.32 & 15.28 & 15.26 & 15.24 \\
\end{tabular}
\label{ganconv}
\end{table}
The agreement of our calculated quasiparticle energies for Si and GaN with
experiment is similar to that of previous {\it GW} calculations but -- in
contrast to these earlier works -- dynamical effects are fully included here.

Fitting of the tails allows to reduce the time/energy 
region which is treated explicitly by more than a factor of three. 
This saves the same factor in the number of FFT grid points. Employing a GL 
grid enables us to reduce the number of points where the functions have to be
computed by another factor between two and three.
In total, the CPU time, memory and disk space requirements decrease by
a factor of seven to eight in comparison with the time/energy FFT grid
treatment described in CPC I. 

\section{Plane wave substitution}
\label{sec:pwt}

\subsection{motivation and basic idea}
\label{sec:pwtmotiv}

A large number of unoccupied states have to be included in the band 
sum in the Green's function Eq.\ (\ref{GLDAST}) for a proper convergence
of the resulting self-energy and QP energies.
With growing system size it becomes increasingly difficult to provide such
a large number of eigenstates by a density-functional calculation since
direct diagonalization may not be computationally feasible whereas
iterative diagonalization yields only a limited number of eigenstates or
becomes prohibitively expensive. Besides that it would be desirable to
accelerate the convergence of the unoccupied-state sum in order to
reduce the computational effort for the calculation of the Green's function.

\begin{figure}
\epsfxsize=3.25in \epsfbox{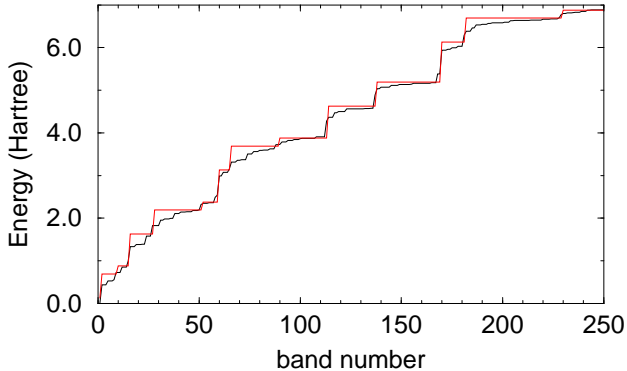}
\caption{Band energies as a function of band number for the LDA eigenstates
of bulk Si (solid line) and the corresponding plane-wave states (dotted
line). The latter have been shifted upwards by 0.26 Ry (with this shift
the energies of LDA and PW state number 50 coincide). At higher energies
the two spectra look remarkably similar (see text). }
\label{ebandpw}
\end{figure}

On the other hand we expect that the higher the energy of an unoccupied 
state the better it should be approximated by a free-electron state (plane 
wave). This is 
illustrated by Figure \ref{ebandpw} showing the band energy as a function
of the band number for (LDA) eigenstates of Si and the corresponding
plane-wave states with wavevectors $\bK = \bk+\bG$, $\bG$ being reciprocal 
lattice vectors
of Si. At higher energies the two spectra look remarkably similar
if we allow for a constant energy shift between the two. Although closer
examination shows that that this assumption is not fully justified for
states with moderately high energies, which, for symmetry reasons, rather 
resemble linear combinations of several plane waves, the sum of all
unoccupied states above a certain energy cutoff can still be reasonably well
described by a corresponding sum of plane waves. The aspect of taking 
proper account
of the weight of the higher unoccupied states seems to be more important
than their explicit form. Thus there is good reason to expect that the number
of unoccupied eigenstates which have to be explicitly included into the
band sum in Eq.\ (\ref{GLDAST}) can be substantially reduced by adding a sum
of plane waves replacing the omitted higher unoccupied states.\cite{foot7}
This is indeed the case as is demonstrated by the significant improvement
of the convergence of the QP energies as a function of the band cutoff
upon adding a plane-wave (PW) contribution to the Green's function,
cf. Figures \ref{si.bconv} and \ref{gan.bconv} below.

\subsection{method}
\label{sec:pwtmethod}

The PW contribution to the Green's function Eq.\ (\ref{GLDAST}) 
takes the following form in real-space:
\begin{eqnarray}
\label{gpw}
\Delta G_{PW}(\br,\br';i\tau) &=& 
   - \frac{i}{V} \sum\limits_{\bK} exp(i\bK\br) exp(-i\bK\br') \nonumber\\
 && \times\; exp(-\frac{\tau}{2}(\bK^2-k_0^2)), 
\end{eqnarray}
where the $\bK$ vectors corresponding to energies below
the lowest PW energy are excluded from the
reciprocal-space sum. The plane waves are normalized with respect to the
crystal volume $V = V_{UC} N_{\bk}$, with $N_{\bk}$ being the number of 
$\bk$ points in the Brillouin zone (BZ) and $V_{UC}$ the 
volume of the unit cell. 
The energies of the PW states are measured with respect to an energy
zero $\Delta E = k_0^2/2$ which is determined by adjusting the 
energy of 
the highest LDA eigenstate included and the highest PW state not 
included in the calculation of the Green's function, see 
Eq.\ (\ref{eshift}) below. $\Delta G_{PW}$ can be computed analytically by
transforming the $\bK$ sum into an integral by 
$\sum_{\bK} \rightarrow V_{UC}/(2\pi)^3 \int d^3K$, i.~e.~taking the
limit $N_{\bK} \rightarrow \infty$ and solving the resulting integral.
It turns out, however, that it is more practical to compute $\Delta G_{PW}$
numerically instead, even though that is computationally slightly more
expensive. In this way the contributions of the (LDA) eigenstates of the 
system and the plane waves are treated on an equal footing which makes for a
smoother convergence because of compensation of errors arising from 
discretization. Fourier transformation of $\Delta G_{PW}(\br,\br';i\tau)$
to reciprocal space results in:
\begin{eqnarray}
\label{gpwk}
\Delta G_{PW}(\bk,\bG,\bG';i\tau) &=& \frac{1}{N_{\bk}V_{UC}} 
exp(-\frac{\tau}{2} k_0^2) \nonumber \\
&& \times\; exp(-\frac{\tau}{2} (\bk+\bG)^2) \delta_{\bG\bG'},
\end{eqnarray}
with reciprocal-lattice vectors $\bG$,$\bG'$ of the system with
$|\bk + \bG|^2/2$ larger than a given cutoff energy.
As Eq.\ (\ref{gpwk}) is diagonal in $\bG$ it is more efficient to first set
up $\Delta G_{PW}$ in reciprocal space and then transform it to real space
before adding it to the Green's function contribution of the LDA eigenstates
taken into account explicitly.

In order not to destroy the crystal symmetry of the Green's function when
adding the PW contribution only complete stars of $\bG$ vectors (groups of
plane-wave states energetically degenerate at any $\bk$ in the BZ) must be
included into or excluded from $\Delta G_{PW}$. On the other hand the number of
$\bG$ excluded from $\Delta G_{PW}$  should be equal to $N_{bands}$, the
number of LDA eigenstates taken into account explicitly, which, in turn, has
to be determined in such a way as to include complete groups of energetically
degenerate (at any $\bk$) LDA eigenstates. 
These two demands cannot in general be fulfilled simultaneously.
As a compromise we calculate $\Delta G_{PW}$ from PW states $(\bk+\bG_i)$ 
(ordered with respect to their energy) with weights
\begin{equation}
\label{weights}
W_i (\bk) = 
         \left\{\begin{array} {lcl}
                    0   &&    i\le N_1(\bk) \\
          1 - \frac{N_{bands} - N_1(\bk)}{N_2(\bk) - N_1(\bk)} & \mbox{for} &
                   N_{1}(\bk) < i \le N_2(\bk) \\
                    1   &&    i > N_2(\bk)
                    \end{array} \right. 
\end{equation}
where $N_1(\bk)$ and $N_2(\bk)$ are the largest possible/smallest 
possible
total number of PW states smaller than/larger than $N_{bands}$, respectively,
both containing complete stars of PW states only.
This $\bk$ dependent cutoff ensures preservation
of both symmetry and number of bands and works well in practice.

In order to account for the difference in the energy spectra of LDA
eigenstates and plane waves we introduce an energy shift
\begin{equation}
\label{eshift}
\Delta E = \frac{1}{2} |G (N_{bands})|^2 - E_{cut} + E_{VBB}
\end{equation}
between the highest PW state(s) not included and the highest LDA eigenstate(s)
included in the band sum in Eq.\ (\ref{GLDAST}). 
$E_{cut}$ and $E_{VBB}$ stand for the energy (at $\bk = 0$) of the highest LDA 
state taken into account and the LDA valence-band bottom, respectively
(both measured with respect to the Fermi level which is chosen halfway 
between valence band top and conduction band bottom)
and $\frac{1}{2} |G (N_{bands})|^2$ is the energy of plane wave number
$N_{bands}$ at $\bk = 0$. Taking this energy shift at $\bk = 0$ 
is somewhat arbitrary, but it turns out
that the resulting self-energies and QP energies are not sensitive to
the exact value of the energy shift.\cite{foot8} 

\begin{figure}
\epsfxsize=3.25in \epsfbox{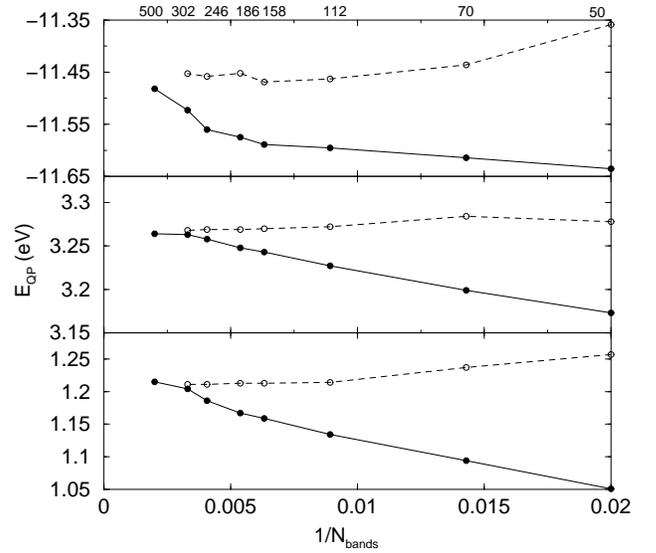}
\caption{Calculated valence band bottom $\Gamma_1^v$ (top panel, the valence 
band maximum has been set to zero), direct gap $\Gamma_{15}^{c\prime}$ (center), 
and minimal gap of bulk Si as a function
of the inverse of $N_{bands}$, the number of LDA eigenstates used for the 
calculation of the Green's function. Two sets of data are shown, the filled
circles (solid lines) refer to calculations with LDA eigenstates only
whereas the open circles (dashed lines) are the results of corresponding
calculations where the PW contribution was added (see text). 
The numbers along the top axis are the $N_{bands}$ values used in the respective
calculations. The number of LDA eigenstates
needed to converge the quasiparticle energy is significantly reduced by 
adding the PW contribution.}
\label{si.bconv}
\end{figure}

\subsection{Tests for bulk Si and GaN}
\label{sec:pwttests}

We performed a number of tests in order to assess the influence of adding
the PW contribution on the convergence of self-energy and QP energies.
The cutoff parameters for our tests for Si are given in Table \ref{pwtpar}.
Further calculational details are as in CPC I. Figure \ref{si.bconv} 
exhibits the convergence of valence band bottom (top panel), direct gap 
at $\Gamma$ (center), and minimal gap (bottom) as a function of the inverse of
$N_{bands}$, the number of LDA eigenstates used for the calculation
of the Green's function. Two sets of data are shown in each figure, showing
results obtained with (open circles) and without (filled circles) adding
the PW contribution. First of all we observe that, as expected, both 
sets of calculations converge to the same answer for 
$1/N_{bands} \rightarrow 0$. However, adding the PW contribution dramatically
improves convergence; the number of eigenstates needed for an accuracy of the
QP energies of 30 meV is about 70 \% smaller when the PW contribution is 
added (see also Table \ref{siresult}.
For the slowly-converging valence band bottom we find that the
number of eigenstates in Eq.\ (\ref{GLDAST}) can be reduced by as much as
85 \%. 

\begin{figure}
\epsfxsize=3.25in \epsfbox{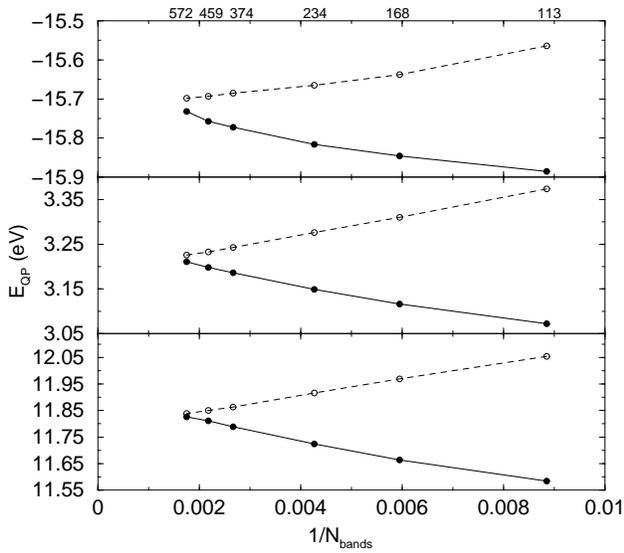}
\caption{Same as figure \ref{si.bconv} for valence band bottom 
$\Gamma_1^v$ (top panel), direct gap
$\Gamma_1^c$ (center), and conduction state $\Gamma_{15}^c$ (bottom)
of zincblende GaN.  }
\label{gan.bconv}
\end{figure}

Since bulk Si might be perceived as a particularly plane-wave like system
we also tested the method for zincblende GaN. 
The cutoff parameters are given in Table \ref{pwtpar}, for further details
of the calculation see Section \ref{sec:testsi} above.
Figure \ref{gan.bconv} showing the 
band-cutoff dependence of valence band bottom (top panel), conduction band 
bottom (center), and conduction state $\Gamma_{15}^c$ (bottom) confirms 
that the PW contribution to the Green's function improves the
band-cutoff convergence in the case of zincblende GaN, too, although not
as much as for bulk Si. Adding the PW contribution decreases the number 
of bands needed to converge the quasiparticle energies of GaN to an 
accuracy of 30 meV by 30 to 50 \% (see also Table \ref{ganresult}).
\begin{table}
\caption{Cutoff and grid parameters used for the test calculations for
Si and zincblende GaN, respectively, in section \ref{sec:pwt} of the present 
work.}
\begin{tabular}{lcc}
 parameter & Si & GaN   \\
\tableline
 LDA plane-wave cutoff (in Ry)                         &  19. & 50. \\
 $GW$ plane-wave cutoff\tablenotemark[1] (in Ry)       &  26. & 50. \\
 $GW$ real-space grid       & $12\times12\times12$  & $15\times15\times15$ \\
 band cutoff\tablenotemark[2] (in Ry)                  &  10. & 27.  \\
 size of {\bf k} grid          & $4\times4\times4$ & $4\times4\times4$   \\
 range $\tau_{max}$ of time grid (in a.u.)             & 6. & 6.  \\
 size of time (energy) grid                            & 15 & 15 \\
\end{tabular}
\label{pwtpar}
\tablenotetext[1]{Energy cutoff corresponding to radius of circumscribing 
sphere, see Ref.\ \onlinecite{RSWRG99}.} 
\tablenotetext[2]{This parameter is varied in the tests of Section 
\ref{sec:pwt}.}
\end{table}
\begin{table}
\caption{Calculated quasiparticle energies at the $\Gamma$ and $X$ point for
Si (in eV) as a function of the number of LDA eigenstates $N_{bands}$ included 
in the calculation of the Green's function. Two sets of results are shown,
obtained with including (b) or not including (a) the PW contribution to the
Green's function (see text). The valence band maximum has been set to zero. }
\begin{tabular}{lrrrrrrr}
 band cutoff [Ry] &    &  6. &   8. & 12.  &  14. & 16. & 18. \\
 $N_{bands}$      &    & 70  & 112  & 186  & 246  & 302 & 500 \\
\tableline
 $\Gamma_1^v$    & (a)& -11.61 & -11.60 & -11.58 & -11.56 & -11.52 & -11.48 \\  
                 & (b)& -11.44 & -11.46 & -11.45 & -11.46 & -11.45 \\[1mm]
 $\Gamma_{15}^{c\prime}$
                 & (a)&   3.20 &   3.23 &   3.25 &   3.26 &   3.26 &   3.26 \\
                 & (b)&   3.28 &   3.27 &   3.27 &   3.27 &   3.27 \\[1mm]
 $\Gamma_2^{c\prime}$   
                 & (a)&   3.93 &   3.94 &   3.95 &   3.96 &   3.96 &   3.96 \\
                 & (b)&   3.92 &   3.94 &   3.95 &   3.96 &   3.96 \\[2mm]
 $X_1^v$         & (a)&  -7.71 &  -7.69 &  -7.68 &  -7.66 &  -7.64 &  -7.63 \\
                 & (b)&  -7.64 &  -7.65 &  -7.64 &  -7.64 &  -7.64 \\[1mm]
 $X_4^v$         & (a)&  -2.85 &  -2.83 &  -2.82 &  -2.81 &  -2.80 &  -2.79 \\
                 & (b)&  -2.78 &  -2.80 &  -2.79 &  -2.80 &  -2.79 \\[1mm]
 $X_1^c$         & (a)&   1.24 &   1.27 &   1.31 &   1.33 &   1.35 &   1.36 \\
                 & (b)&   1.38 &   1.36 &   1.36 &   1.35 &   1.35 \\[1mm]
 $X_4^c$         & (a)&  10.75 &  10.75 &  10.75 &  10.74 &  10.72 &  10.71 \\
                 & (b)&  10.67 &  10.70 &  10.70 &  10.70 &  10.70 \\
\end{tabular}
\label{siresult}
\end{table}
\begin{table}
\caption{Same as Table \ref{siresult} for zincblende GaN.}
\begin{tabular}{lrrrrrrr}
 band cutoff [Ry] &    & 12. &  16. & 20.  &  28. & 32. & 36. \\
 $N_{bands}$      &    & 113 & 168  & 234  & 374  & 459 & 572 \\
\tableline
 $\Gamma_1^v$    & (a)& -15.88 & -15.85 & -15.82 & -15.77 & -15.76 & -15.73 \\  
                 & (b)& -15.56 & -15.64 & -15.66 & -15.68 & -15.69 & -15.70 \\[1mm]
 $\Gamma_1^c$    & (a)&   3.07 &   3.12 &   3.15 &   3.19 &   3.20 &   3.21 \\
                 & (b)&   3.38 &   3.31 &   3.28 &   3.24 &   3.23 &   3.23 \\[1mm]
 $\Gamma_{15}^c$ & (a)&  11.58 &  11.66 &  11.72 &  11.79 &  11.81 &  11.83 \\
                 & (b)&  12.05 &  11.97 &  11.92 &  11.86 &  11.85 &  11.84 \\[2mm]
 $X_1^v$         & (a)& -12.93 & -12.91 & -12.90 & -12.87 & -12.86 & -12.85 \\
                 & (b)& -12.75 & -12.81 & -12.82 & -12.83 & -12.84 & -12.84 \\[1mm]
 $X_3^v$         & (a)&  -6.39 &  -6.34 &  -6.30 &  -6.26 &  -6.24 &  -6.22 \\
                 & (b)&  -6.10 &  -6.15 &  -6.17 &  -6.20 &  -6.20 &  -6.21 \\[1mm]
 $X_5^v$         & (a)&  -2.66 &  -2.63 &  -2.61 &  -2.59 &  -2.58 &  -2.57 \\
                 & (b)&  -2.52 &  -2.54 &  -2.55 &  -2.56 &  -2.57 &  -2.57 \\[1mm]
 $X_1^c$         & (a)&   4.46 &   4.55 &   4.61 &   4.67 &   4.70 &   4.72 \\
                 & (b)&   4.94 &   4.86 &   4.81 &   4.76 &   4.75 &   4.74 \\[1mm]
 $X_1^{c\prime}$ & (a)&   7.94 &   8.01 &   8.06 &   8.11 &   8.13 &   8.15 \\
                 & (b)&   8.33 &   8.26 &   8.22 &   8.18 &   8.17 &   8.16 \\
 $X_3^c$         & (a)&  13.27 &  13.28 &  13.29 &  13.30 &  13.31 &  13.31 \\
                 & (b)&  13.31 &  13.31 &  13.31 &  13.31 &  13.31 &  13.31 \\
 $X_5^c$         & (a)&  15.35 &  15.38 &  15.40 &  15.42 &  15.42 &  15.43 \\
                 & (b)&  15.46 &  15.45 &  15.44 &  15.43 &  15.43 &  15.43 \\
\end{tabular}
\label{ganresult}
\end{table}

\begin{figure}
\epsfxsize=3.25in \epsfbox{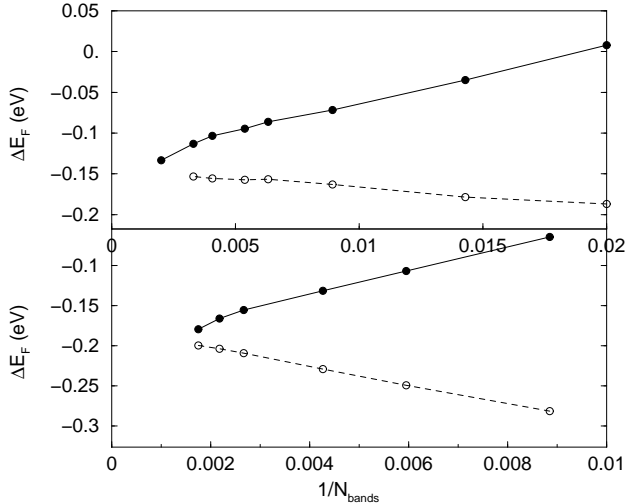}
\caption{Fermi energy shift $\Delta E_F = E_F^{QP}-E_F^{LDA}$ for bulk Si
(top panel) and zincblende GaN (bottom panel) as a function
of the inverse of $N_{bands}$, the number of LDA eigenstates used for the 
calculation of the Green's function, calculated including (open circles,
dashed lines) or not including (filled circles, solid lines) the plane-wave
contribution (see text). The plane-wave contribution improves the convergence 
of absolute self-energies in qualitatively the same way as that of the QP 
energy differences shown in figures \ref{si.bconv} and \ref{gan.bconv}.
}
\label{efshift}
\end{figure}

It can be concluded from the Fermi energy shifts 
$\Delta E_F = E_F^{QP} - E_F^{LDA}$ 
shown in Figure \ref{efshift} that adding the PW contribution improves the 
convergence of absolute self-energies in qualitatively the same way as that of 
QP energy differences (Figures \ref{si.bconv} and \ref{gan.bconv}).

In summary, we find that the number of LDA eigenstates (bands)
needed to converge the QP energies within 30 meV can be considerably
reduced by including the PW contribution described in Section 
\ref{sec:pwtmethod} in the calculation of the Green's function.

\section{Summary}

In the present work we described two new features which significantly 
enhance the power of the real-space imaginary-time $GW$ scheme
for the calculation of self-energies and related quantities of solids.
Fitting the smoothly decaying large-imaginary-energy/time tails and
treating the remaining imaginary energy/time region numerically on a 
Gauss-Legendre grid allows to reduce the computational time and storage 
requirements of the method by a factor of seven to eight while retaining the 
flexibility to accomodate general functional forms of the energy 
dependence.\cite{foot9}
The tail-fitting procedure suggested in the present work turned out to be 
accurate and reliable.
Substituting the contribution of higher unoccupied eigenstates to the Green's
function Eq.\ (\ref{GLDAST}) with a sum of corresponding free-electron states
(plane waves) accelerates the convergence of the eigenstate sum in
Eq.\ (\ref{GLDAST}), thus substantially reducing the number of eigenstates 
and eigenvalues which have to be provided by a density-functional calculation 
preceding the calculation of the self-energy and simultaneously decreasing
the computational effort for the calculation of the Green's function itself.

\section{Acknowledgments}

This work was supported by the Engineering and Physical Sciences Research 
Council, the Spain-UK Acciones Integradas program (HB 1997-011), 
JCyL (Grant: VA28/99) and DGES (Grant: PB95-0720). 
L. Reining acknowledges a grant of computer time on the C98 of IDRIS 
(project CP9/980544), which was used for parts of the calculations.


\begin{thebibliography}{10}
\bibitem{foot1}
If we were working on the real energy axis this number would be an order of
  magnitude larger still, as can be concluded from Ref.\
  \onlinecite{Fleszar97}.
\bibitem{foot3}
An alternative to a GL grid would be a grid which is more dense towards the
  origin by a suitable transformation of variables. However, such grids do not
  perform well in computing the `fast-oscillating' components, i.~e.~$i\int
  d\tau f(i\tau) exp(-i\omega\tau)$ for large $\omega$.
\bibitem{foot4}
In our experience fitting in reciprocal space rather than in real space is both
easier and more efficient. 
For very large systems where it is advantageous to use a real-space representation
throughout\cite{RSWRG99} the fit can of course also be done in real space.
\bibitem{foot5}
taking the free-electron-gas plasmon energy at the average valence density of
  bulk Si that corresponds to between 2 and 6 Hartree.
\bibitem{foot8}
For example, setting this energy shift to zero changes the calculated QP
  energies by less than 10 meV in the case of bulk Si. Of course the QP
  energies always converge to the same values as in a calculation where no
  plane waves were added since the plane wave contribution vanishes for
  $N_{bands} \rightarrow \infty$.
\bibitem{foot9}
The CPU time (on a Digital Alpha 500/500 workstation), disk space and memory 
required for computing the full self-energy and 48 self-energy matrix elements 
for Si (GaN) are 35 minutes (194 minutes), 243 MB (1213 MB) and 65 MB (123 MB),
respectively, with the parameters given in Table \ref{glgpar}.
\end{thebibliography}



\end{document}